\documentclass[%
 aip,
 amsmath,amssymb,
 reprint,%
]{revtex4-2}

\usepackage[version=3]{mhchem} 
\usepackage{xcolor}
\usepackage{siunitx}
\usepackage{amssymb}

\usepackage{graphicx}
\usepackage{dcolumn} 
\usepackage{bm}
\usepackage[utf8]{inputenc}
\usepackage[T1]{fontenc}
\usepackage{mathptmx}
\usepackage{etoolbox}
\usepackage{relsize}

\usepackage{xr}
\newcommand{\Pp}{P_{\mathrm{p}}}
\newcommand{\Pe}{P_{\mathrm{e}}}
\newcommand{\Pd}{P_{\mathrm{d}}}
\newcommand{\Psat}{P_{\mathrm{sat}}}
\newcommand{\lesssimsmall}{\,\mathsmaller{\lesssim}\,}

\makeatletter
\def\@email#1#2{%
 \endgroup
 \patchcmd{\titleblock@produce}
  {\frontmatter@RRAPformat}
  {\frontmatter@RRAPformat{\produce@RRAP{*#1\href{mailto:#2}{#2}}}\frontmatter@RRAPformat}
  {}{}
}%
\makeatother

\begin{document}
\preprint{AIP/123-QED}

\title{High-resolution vibronic spectroscopy of a single molecule embedded in a crystal} 

\author{Johannes Zirkelbach}
\affiliation{Max Planck Institute for the Science of Light, 91058 Erlangen, Germany}
\affiliation{Department of Physics, Friedrich-Alexander University Erlangen-N\"urnberg, 91058 Erlangen, Germany}

\author{Masoud Mirzaei}
\affiliation{Max Planck Institute for the Science of Light, 91058 Erlangen, Germany}
\affiliation{Department of Physics, Friedrich-Alexander University Erlangen-N\"urnberg, 91058 Erlangen, Germany}

\author{Irena Deperasi\'{n}ska}
\affiliation{Institute of Physics, Polish Academy of Sciences, Al.\,Lotnik\'{o}w 32/46, 02-668 Warsaw, Poland}

\author{Boleslaw Kozankiewicz}
\affiliation{Institute of Physics, Polish Academy of Sciences, Al.\,Lotnik\'{o}w 32/46, 02-668 Warsaw, Poland}

\author{Burak Gurlek}
\affiliation{Max Planck Institute for the Science of Light, 91058 Erlangen, Germany}
\affiliation{Department of Physics, Friedrich-Alexander University Erlangen-N\"urnberg, 91058 Erlangen, Germany}

\author{Alexey Shkarin}
\affiliation{Max Planck Institute for the Science of Light, 91058 Erlangen, Germany}

\author{Tobias Utikal}
\affiliation{Max Planck Institute for the Science of Light, 91058 Erlangen, Germany}

\author{Stephan Götzinger}
\affiliation{Max Planck Institute for the Science of Light, 91058 Erlangen, Germany}
\affiliation{Department of Physics, Friedrich-Alexander University Erlangen-N\"urnberg, 91058 Erlangen, Germany}
\affiliation{Graduate School in Advanced Optical Technologies (SAOT), Friedrich-Alexander
University Erlangen-Nuremberg, 91052 Erlangen, Germany}

\author{Vahid Sandoghdar}
\email{vahid.sandoghdar@mpl.mpg.de}	
\affiliation{Max Planck Institute for the Science of Light, 91058 Erlangen, Germany}
\affiliation{Department of Physics, Friedrich-Alexander University Erlangen-N\"urnberg, 91058 Erlangen, Germany}

\begin{abstract}
Vibrational levels of the electronic ground states in dye molecules have not been previously explored at high resolution in solid matrices. We present new spectroscopic measurements on single polycyclic aromatic molecules of dibenzoterrylene embedded in an organic crystal made of para-dichlorobenzene. 
To do this, we use narrow-band continuous-wave lasers and combine spectroscopy methods based on fluorescence excitation and stimulated emission depletion (STED) to assess individual vibrational linewidths in the electronic ground state at a resolution of $\sim$30 MHz dictated by the linewidth of the electronic excited state.
In this fashion, we identify several exceptionally narrow vibronic levels with linewidths down to values around $\SI{2}{GHz}$. Additionally, we sample the distribution of vibronic wavenumbers, relaxation rates, and Franck-Condon factors, both in the electronic ground and excited states for a handful of individual molecules. We discuss various noteworthy experimental findings and compare them with the outcome of DFT calculations. The highly detailed vibronic spectra obtained in our work pave the way for studying the nanoscopic local environment of single molecules. The approach also provides an improved understanding of the vibrational relaxation mechanisms in the electronic ground state, which may help to create long-lived vibrational states for applications in quantum technology.
\end{abstract}

\maketitle

\section*{Introduction}
Molecules are formed when at least two atoms are chemically bound and, as such, molecular science is often primarily associated with the realm of chemistry. Of course, molecules are also used for studies in physics, e.g., for harnessing the solar energy or use in fluorescence microscopy. However, the fact that molecules provide some of the most intricate quantum systems has not been widely acknowledged among physicists. With the rapid emergence of quantum technologies, this is changing, and more researchers are beginning to turn their attention to the mechanical properties of molecules for their potential as quantum optomechanical oscillators or as qubits \cite{tesch-2002, suzuki-2005, roelli-2016, gurlek-2021}. It is highly likely that this line of quantum technology work will heavily rely on single-molecule measurements.

Optical detection of single quantum emitters was a holy grail in physical sciences during the 1970s and 1980s. Experiments on single trapped ions in an ultrahigh vacuum chamber and the observation of quantum jumps established a sensational start, but the largest impact came from the detection of single molecules in the solid state as pioneered by W.E. Moerner in 1989 and Michel Orrit in 1990.\cite{moerner-1989,orrit-1990} The first demonstrations were performed in absorption mode at liquid helium temperature on the organic dye molecule pentacene embedded in the organic crystal para-terphenyl. Orrit and Bernard presented results on the same system but via the detection of fluorescence photons on a dark background, thus, yielding a larger signal-to-noise ratio (SNR).  In the decade that followed, single-molecule detection rapidly spread to a very wide range of modalities (frequency selection, spatial selection, near field, confocal, wide field, etc), media (fluids, solids, crystals, polymers, etc.) and applications (physical chemistry, physics, biology, materials science, etc) \cite{kozankiewicz-2014, orrit-2014}. One of the elegant consequences of single-molecule science has been the emergence of molecular quantum optics, which has reached the same level of sophistication that is known from experiments on gaseous atoms, GaAs-based quantum dots, or color centers in diamond over the past two decades \cite{hwang-2009, maser-2016, rattenbacher-2019, wang-2019, zirkelbach-2020, pscherer-2021}. These studies
take place at liquid helium temperature, where the zero-phonon line (ZPL) transition reaches its Fourier limit in the order of 10-50\,MHz. The foundation of molecular quantum optics was set in the 1990s by Orrit, Lounis and coworkers through a series of beautiful experiments that established single molecules as well-defined emitters for basic quantum optical studies and as bright sources of single photons \cite{orrit-1992, basche-1992, tamarat-1995, lounis-1997, brunel-1999, lounis-2005}. Indeed, it was many of these single-molecule developments that were soon adopted by a wide community of researchers for the detection of other species such as semiconductor quantum dots, color centers, nanotubes, ions, etc.~\cite{gammon-1996, gruber-1997, hartschuh-2003, utikal-2014}.

One of the desirable features of quantum systems is long coherence times. Considering that the excited states of optical emitters have lifetimes in the order of nanoseconds, many groups have explored the use of spin states in atoms, semiconductor quantum dots, color centers, or rare earth ions in crystals for longer lifetimes \cite{specht-2011, wang-2021, muhonen-2014, hanson-2007, herbschleb-2019, siyushev-2014}. Stable dye molecules do not offer spin degrees of freedom in their ground and excited electronic states because electrons are usually paired, resulting in singlet states. An intriguing possibility for accessing long lifetimes in molecules might, however, be to exploit their vibrational states. Indeed, the lifetime of the stretch mode of N$_2$ molecules in liquid nitrogen appears to be \SI{56}{s} governed by infrared spontaneous emission \cite{brueck-1976, everitt-2002}. Similarly, the vibrations of some other diatomic molecules in traps and supersonic jets were found to be on time scales of the order of \SI{10}{ms}, also limited by radiative decay \cite{drabbels-1997, jongma-1997, meerakker-2005, campbell-2008}. In polyatomic molecules, intramolecular vibrational redistribution (IVR) of population to an internal manifold of lower-lying vibrational states imposes an additional fundamental constraint on lifetimes, often reducing them to values below \SI{1}{ns}.\cite{lehmann-1994, nesbitt-1996}. 

The relaxation rate of molecular vibrations increases further if the molecule couples to the mechanical degrees of freedom in its environment. In the gas phase, vibrational energy can be released by collisions with other molecules. In the condensed phase, phonon modes of the matrix and other modes of the environment can carry the excess energy away from the molecule. In this case, typical vibrational lifetimes of (polyatomic) molecules are lowered to values between 1--\SI{100}{ps}, depending on the molecular structure, the vibrational mode, and its mechanical coupling to the environment.\cite{dlott-1989, elsaesser-1991} 

Recent developments in cooling and trapping of molecules have enabled access to single molecules in vacuum.\cite{anderegg-2019} 
Vibrational levels are expected to have low decoherence rates at these conditions, but the excitation of long-lived vibrations might be hampered by the optical cycling required for laser cooling. In a quest towards the use of molecular vibrational states for applications in quantum information processing, it is important to explore the practical limits of population and coherence lifetimes in solid-state systems. 
In particular, one is prompted to search for systems with low IVR rates and good isolation from the environment \cite{tesch-2002}. In this work, we employ laser spectroscopy on single molecules to examine the linewidths of the various vibrational states both in the electronic excited and ground states at $T\lesssimsmall100$\,mK. Our efforts set the ground for investigating the effects of external interactions for controlling various vibrational lifetimes and coherences.

\section*{The molecular system}
\begin{figure*}
  \includegraphics{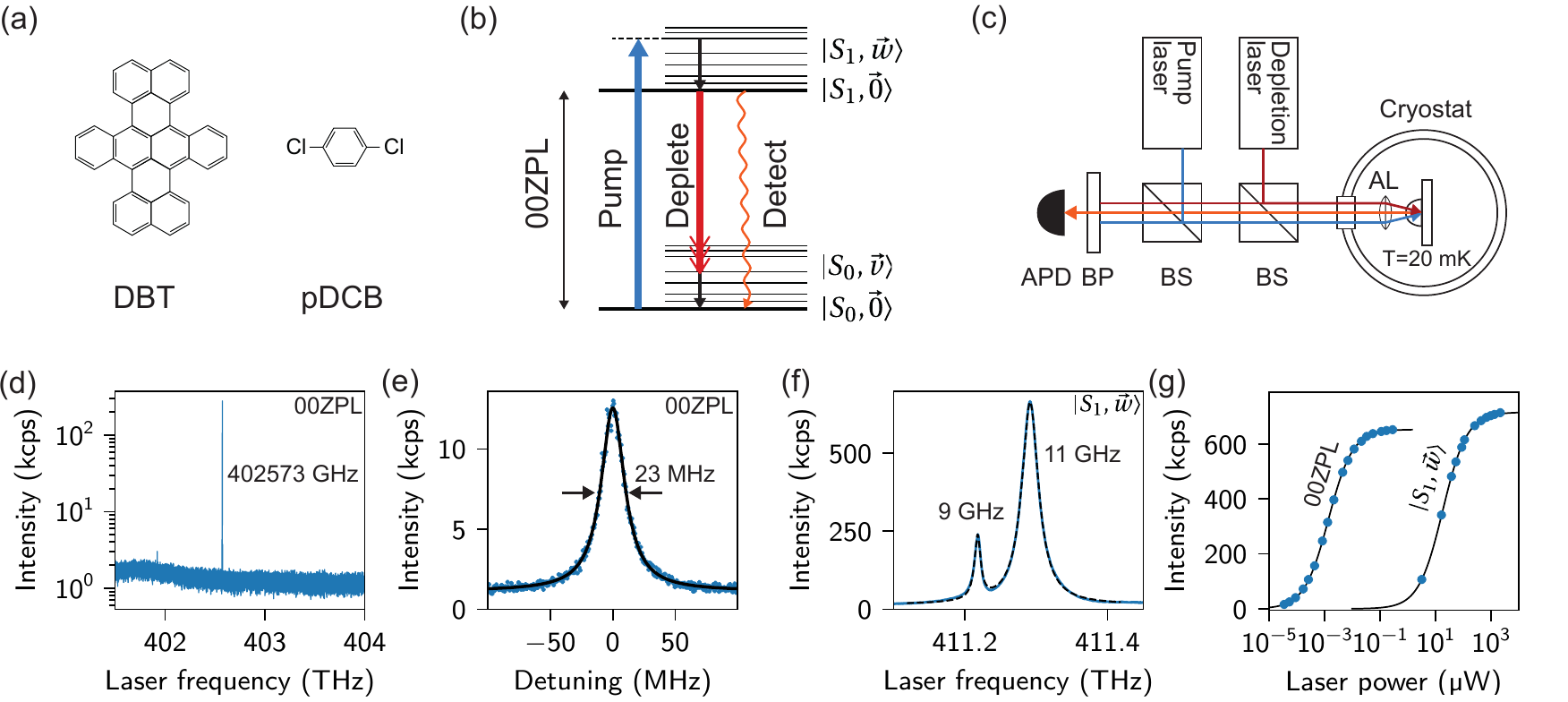}
  \caption{Molecular system and experimental setup. 
  (a) Molecular structures of dibenzoterrylene (DBT) and para-dichlorobenzene (pDCB).
  (b) Schematic level scheme of DBT. The relevant transitions addressed by the pump (excitation) and depletion lasers as well as the fluorescence decay are marked by blue, maroon and orange arrows, respectively. Multi-headed arrows indicate frequency scanning. Black arrows indicate fast vibrational relaxation.
  $\lvert S_0, \vec{v} \rangle$: vibronic levels associated to the electronic ground state.
  $\lvert S_1, \vec{w} \rangle$: vibronic levels associated to the electronic excited state.
  (c) Simplified sketch of the optical setup used for stimulated emission depletion spectroscopy.
  If the depletion laser is blocked, the setup can be used for fluorescence excitation spectroscopy.
AL: aspheric lens (NA: 0.7), BS: beam splitter, BP: bandpass, APD: avalanche photodiode.
  (d) Fluorescence excitation scan showing the isolated 00ZPL resonance of a single DBT molecule in a spectral range of more than 2\,THz within the typical inhomogeneous spectrum of DBT:pDCB ($\Pe = \SI{3}{nW}$).
  (e) Fluorescence excitation spectrum around the 00ZPL shown in (d) recorded at a low excitation power ($\Pe = \SI{0.03}{nW}$). Black line indicates a Lorentzian fit, revealing a linewidth of about 23\,MHz.
  (f) Fluorescence excitation scan of the most prominent vibronic transition in $S_1$ and a weak additional vibronic level at lower frequency ($\Pe = \SI{201}{\micro W}$). 
  The black dashed line presents the fit obtained from a rate equation model, resulting in the indicated linewidths.
(g) Saturation curves recorded under excitation via 00ZPL ($\Psat = \SI{1.4}{nW} $) and via the prominent vibronic transition shown in (f) ($\Psat = \SI{18.6}{\micro W}$ for the prominent transition).
Black lines are fits of the function $R_{\infty} \cdot S_{\mathrm{e}} / (1+ S_{\mathrm{e}})$ with $S_{\mathrm{e}} = \Pe / \Psat$.
The values of $R_{\infty}$ are not comparable in this plot because the spectral regions used for detection were different between the two measurements.
We estimate that a laser power of  \SI{1}{nW} in front of the cryostat translates to an intensity of $\lesssimsmall$\SI{0.5}{W/cm^2} at the position of the molecules.
Differences in focus quality and the position of the molecules relative to the optical axis of the SIL may cause variations in the local intensity.}
  \label{schematic}
\end{figure*}

In this work, we record high-resolution vibronic spectra of dibenzoterrylene (DBT; C$_{38}$H$_{20}$) molecules in a para-dichlorobenzene (pDCB) crystal (see Figure \ref{schematic}(a)). The potential surfaces of the electronic ground state $\vert S_0 \rangle$ and excited state $\vert S_1 \rangle$ of DBT have a small displacement with respect to each other. As a consequence, the ZPL transition between the vibrational ground states $\lvert S_0, \vec{v}=\vec{0} \rangle$ and $\lvert S_1, \vec{w}=\vec{0} \rangle$ (00ZPL) has a large Franck-Condon (FC) overlap (see Figure \ref{schematic}(b) and Section I of the Supplementary Material (SM) for notation). The absence of thermal phonons at cryogenic temperatures leads to an environment with negligible electronic dephasing. This results in Fourier-limited linewidths in the order of 10-50\,MHz, giving rise to an efficient interaction with light. Slight variations in the local solid matrix cause the transition frequencies of the individual molecules in a sample to differ from each other. It was, indeed, the resulting inhomogeneous distribution of resonance frequencies that allowed scientists to address \textit{single} molecules via spectral selection of their narrow 00ZPLs \cite{basche-2008}. An important advantage of solid-state molecules over those in the liquid and gas phases is that translational and rotational degrees of freedom are suppressed, tremendously simplifying their level structure and making it possible to study a single molecule at the same position over a period of months.

In molecular crystals, the vibronic states $\lvert S_1, \vec{w}\neq \vec{0} \rangle$ are known to decay within typical lifetimes in the range of \SI{10}{ps} to the $\lvert S_1, \vec{w}=\vec{0} \rangle$ state.\cite{hill-1988, plakhotnik-2002} Such fast relaxations broaden the transitions to vibronic states by a factor of 100--1000 as compared to the 00ZPL transition at cryogenic temperatures. If a low enough doping level is used to avoid spectral overlap of the vibronic transitions among nearby molecules, it is possible to excite single molecules selectively also via transitions to the vibronic states of $S_1$. \cite{banasiewicz-2007, wiacek-2008} At $T \lesssimsmall \SI{2}{K}$, the linewidths of these transitions are expected to be determined by vibrational relaxation. The two most important thermal broadening mechanisms, namely phonon-induced dephasing of the electronic transition and an increase of the vibrational relaxation rate caused by thermal phonons, can safely be ignored at these temperatures.\cite{dlott-1989}

The symmetrical behavior of the absorption and emission processes suggests that the transitions $\lvert S_0, \vec{v}=\vec{0} \rangle \rightarrow \lvert S_1, \vec{w} \rangle$ have a counterpart $\lvert S_1, \vec{w}=\vec{0} \rangle \rightarrow \lvert S_0, \vec{v} \rangle$. The latter vibrational states could be of particular interest for quantum information processing because they belong to the electronic ground state and one could hope to establish conditions for reaching long lifetimes. However, investigations of the linewidths of vibronic transitions in single molecules have remained scarce and limited to the vibrational levels of the electronic excited state via $\vert S_0, \vec{0} \rangle \rightarrow \vert S_1 \vec{w} \rangle$ excitation. \cite{nonn-2001, plakhotnik-2002} 
In this study, we employ stimulated emission depletion (STED)\cite{hell-1994, blom-2017} spectroscopy to probe $\lvert S_1, \vec{w}=\vec{0} \rangle \rightarrow \lvert S_0, \vec{v} \rangle$ transitions in single molecules at high spectral resolution. The technique is similar to stimulated emission pumping, which has been used for spectroscopy and vibrational state preparation of molecules in the gas phase\cite{kittrell-1981, hamilton-1986}. In this manner, we improve the spectral resolution by two orders of magnitude with respect to the conventional studies using grating spectrometers.\cite{nicolet-2007, verhart-2016}


 With 58 atoms, DBT has $N_{\mathrm{vib}} = 3\cdot58 - 6 = 168$ vibrational normal modes. Its most stable stereoisomer in vacuum is non-planar and part of the $C_{2h}$ group, both in $S_0$ and $S_1$ states.\cite{deperasinska-2010} Similar bond lengths and nuclear positions in $S_0$ and $S_1$ lead to a strong 00ZPL. In isolated DBT, 41 of the normal modes are totally symmetric (A$_{\mathrm{g}}$). Ab initio calculations predict small Duschinsky mixing\cite{duschinsky-1937} of most normal modes between electronic ground and excited states such that one can define pairs of corresponding modes in both states. In the non-mixing approximation, calculations reveal that the normal coordinate shift $\Delta Q_i$ between the vibrational wavefunctions of $S_0$ and $S_1$ is smaller than the width of the wavefunction in the ground state of the $i$-th mode. Consequently, vibronic transitions from the vibrational ground state to high overtones or combination modes have low FC overlaps. As a result, fluorescence excitation and emission spectra of DBT are dominated by fundamental excitations of totally symmetric normal modes with the highest FC activity. 
 
When embedded in a crystal, the spatial conformation of a molecule is usually distorted. This affects the frequencies of the molecular vibrations and the intensity distribution of the vibronic transitions, which  are either reshuffled between already active lines or partially transferred to new, previously forbidden modes.\cite{myers-1994, kulzer-1997}
For terrylene molecules, the appearance of several strong new lines could be explained by a non-planar deformation in various crystals.\cite{deperasinska-2017, bialkowska-2017}
In the case of DBT, however, the experimental observation of three characteristic strong lines in the $<\SI{300}{cm^{-1}}$ wavenumber region \cite{nicolet-2007, deperasinska-2011, makarewicz-2012, verhart-2016} as well as simulations in anthracene\cite{nicolet-2007-2, makarewicz-2012}, dimethylnaphtphalene\cite{deperasinska-2011}, and dibromonaphthalene\cite{moradi-2019} indicate that DBT does not change its conformation drastically as compared to its isolated case. A quantitative calculation of how embedding DBT in a matrix affects the FC factors of its vibrational modes has not been previously reported and will be presented for pDCB later in this article.

For our spectroscopic measurements, we prepared pDCB crystals with a 30 ppb molar concentration of DBT guest molecules (see Section II of SM). The molten crystal was introduced into channels of about \SI{500}{nm} depth, formed between etched structures on a SiO$_2$ substrate and a solid immersion lens (SIL; $n=2.14$) that covered the top side \cite{zirkelbach-2020}. When the crystal is solidified in the channel, molecules take on random spatial positions in the matrix. At the low doping level used here, we found on average fewer than one molecule per excitation volume of the laser beam, allowing for spatial selection of single molecules within the diffraction limit. To ensure removing thermally induced dephasing from the crystal and to obtain sharp spectral lines, we cooled the sample down to $T\sim \SI{20}{mK}$ in a dilution refrigerator (see Figure \ref{schematic}(c) and Section III of SM). We remark, however, that the local temperature of the sample might be higher in some measurement runs under stronger excitation. 


\section*{Laser spectroscopy}
\subsection*{Fluorescence excitation spectroscopy}
Figure \ref{schematic}(d) shows a fluorescence excitation frequency scan of a molecule at a wavelength of about $\SI{744.7}{nm}$, close to the center of the inhomogeneous broadening that has been reported for DBT in pDCB samples at higher concentrations.\cite{verhart-2016} To obtain this spectrum, the frequency of a narrow-band, tunable continuous-wave Ti:Sapphire laser was scanned through the 00ZPL and the fluorescence generated upon the decay of the excited state was registered. The lack of any other spectral features of appreciable intensity over a range of more than \SI{2}{THz}, which is a much larger span than the reported inhomogeneous broadening of about \SI{120}{GHz} for bulk samples \cite{verhart-2016}, lets us safely assume that there is only one DBT molecule in the excitation volume of the laser beam. 
Furthermore, the 00ZPL linewidth of \SI{23}{MHz} at low laser powers (see Figure \ref{schematic}(e)) agrees well with the lifetime-limited expected value $\Delta \nu = 1/(2\pi T_1)$ determined from pulsed measurements of the excited state lifetime $T_1 = \SI{7}{ns}$ (see SM, section IV). 

\begin{figure*}
  \includegraphics{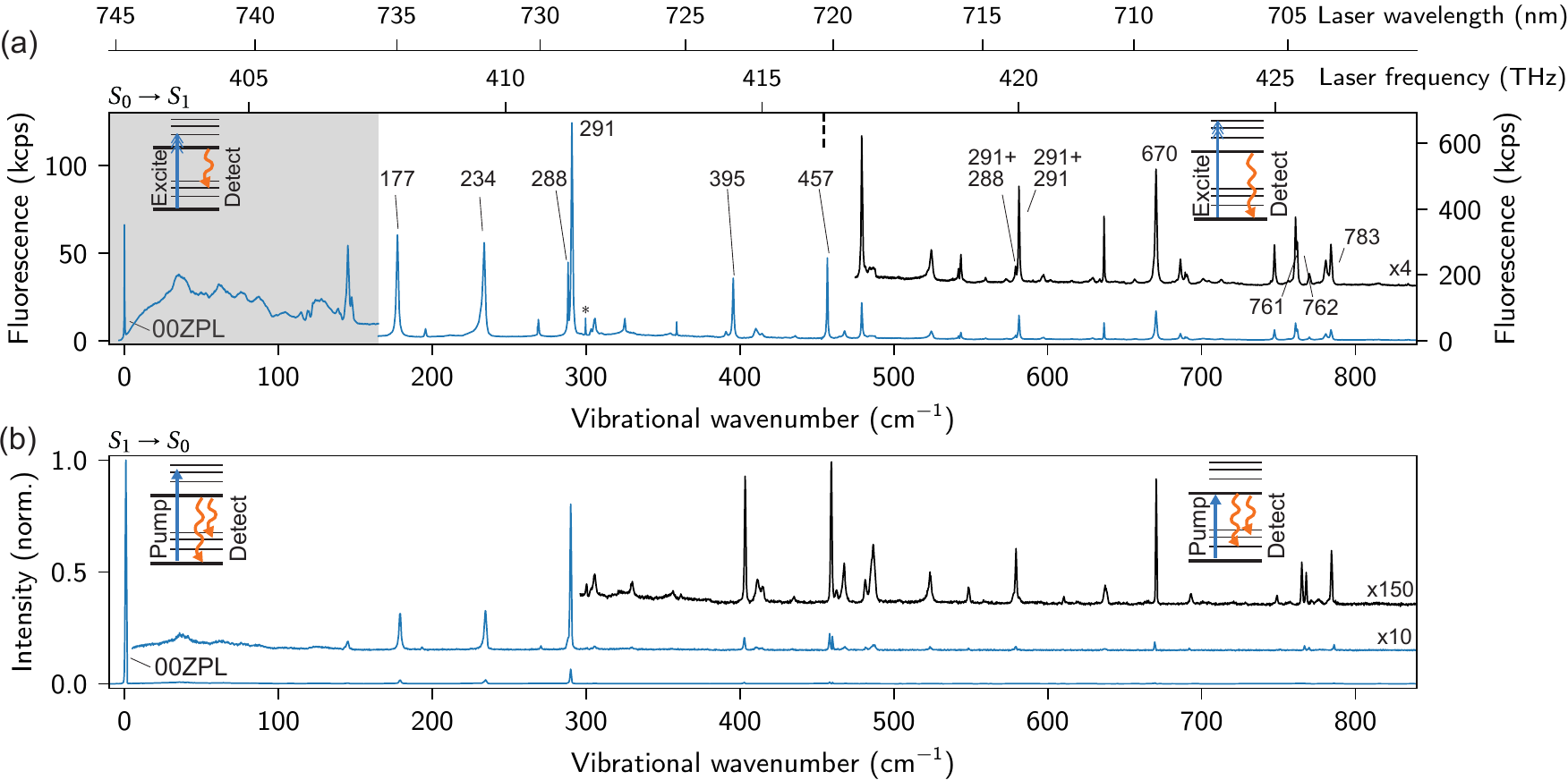}
  \caption{(a) Fluorescence excitation scan of the phonon wing (gray background, left intensity axis) and vibrational states in $S_1$ (white background, right intensity axis). 
  The configurations of excitation ($\Pe = \SI{201}{\micro W}$) and detection are indicated in the plot. Multi-head arrows indicate frequency scanning.
  The vertical black dashed line shows the division between the two scan regions for which the focus of the laser was optimized separately. The inset shows a four-fold magnification of the upper wavenumber region. The numbers in the plot mark selected vibronic transitions analyzed in this work. The region with narrow vibronic states shown in Figure \ref{narrow_lines} is indicated by an asterisk.
  (b) Fluorescence emission spectrum recorded with a grating spectrometer. The blue curves were recorded when exciting the molecule via a higher vibrational level. The upper curve shows a four-fold magnification. The black spectrum was recorded while exciting the molecule via its 00ZPL and corresponds to a 150-fold magnification of the signal.}
  \label{spectra_conventional}
\end{figure*}

We also used fluorescence excitation spectroscopy to probe the vibrational levels of $S_1$. Here, we tuned the laser frequency from the 00ZPL towards higher frequencies while detecting red-shifted fluorescence. Two vibronic peaks recorded in that fashion are shown in Figure \ref{schematic}(f). Due to the fast relaxation of the vibrational states and the reduced FC factors of vibronic transitions, vibronic excitation requires considerably more laser power than pumping via the 00ZPL. As shown in Figure \ref{schematic}(g), we needed about $10^4$ fold more power to saturate the most prominent vibronic transition in $S_1$ at around \SI{299}{cm^{-1}} than for the 00ZPL. We note that a key factor in efficient excitation of the vibrational states has been the use of a SIL, which confines the excitation volume due to its high refractive index ($n$ = 2.14), thus, reducing the saturation power of vibronic transitions to the sub-\si{mW} level \cite{wrigge-2008}. 

Single-molecule spectroscopy requires that only one molecule of the desired species lies in the excitation volume and within the given spectral bandwidth. As displayed in Figure\,\ref{spectra_conventional}(a), we were able to excite the phonon wing and many vibronic states of a single DBT molecule in $S_1$ by choosing sufficiently low doping levels during sample preparation (see also Figure S8(a) of the SM).
By using a suitable arrangement of (tunable) filters in the detection path, we either collected the red-shifted fluorescence while scanning through the phonon wing (gray background) or the light emitted on the 00ZPL for the domain starting from \SI{165}{cm^{-1}} (white background). The three prominent lines at \SI{177}{cm^{-1}}, \SI{234}{cm^{-1}}, and \SI{290}{cm^{-1}} are clearly visible in this spectrum, as expected for the most stable isomer of DBT, together with many additional vibronic features of lower amplitudes.\cite{makarewicz-2012} We note that the range of this spectrum is considerably larger than the fluorescence excitation spectra previously reported in the literature \cite{nonn-2001, plakhotnik-2002, wiacek-2008, deperasinska-2011, bialkowska-2017} 
and is only limited by the tuning range ($\lambda \geq \SI{700}{nm}$) of our Ti:Sapphire laser. 

We remark that to compensate for chromatic aberrations of the aspheric lens used as microscope objective in the cryostat, we split the laser frequency scan between \SI{165}{cm^{-1}} and \SI{800}{cm^{-1}} into two parts for which the collimation of the laser beam was optimized. The black dashed line on the top of the plot indicates the separation between these regions. 

To determine the wavenumbers, linewidths, and relative intensities of the vibronic features in the spectra, we fitted the data with a rate equation model for the excited state population. We start by estimating the (normalized) excited state population by comparing the count rate $R$ recorded on the APD with the saturation count rate $R_{\infty}$ determined in a separate measurement (see Figure \ref{schematic}(g)). The number of vibronic states included in the model was chosen empirically, based on the number of peaks in the data. The fit of the decay rates takes into account power-broadening and directly yields the lifetime-limited linewidth $\Gamma_{we}/(2\pi)$ of the vibrational levels accounted for in the model. For the large vibronic peak of Figure \ref{schematic}(f), we obtain a natural linewidth of $\Gamma_{we} / (2\pi) =\SI{10.9}{GHz}$, corresponding to a vibrational lifetime of $\tau_w = \SI{14}{ps}$. As discussed in more detail below, we find a considerable variation among the observed linewidths between the vibronic states. By comparing spectra between several molecules, we can also quantify typical variations of the vibronic modes among single molecules and between $S_0$ and $S_1$ states (see Figure \ref{stats}).


\subsection*{Grating spectroscopy of vibrational levels in $S_0$}
\begin{figure*}
  \includegraphics{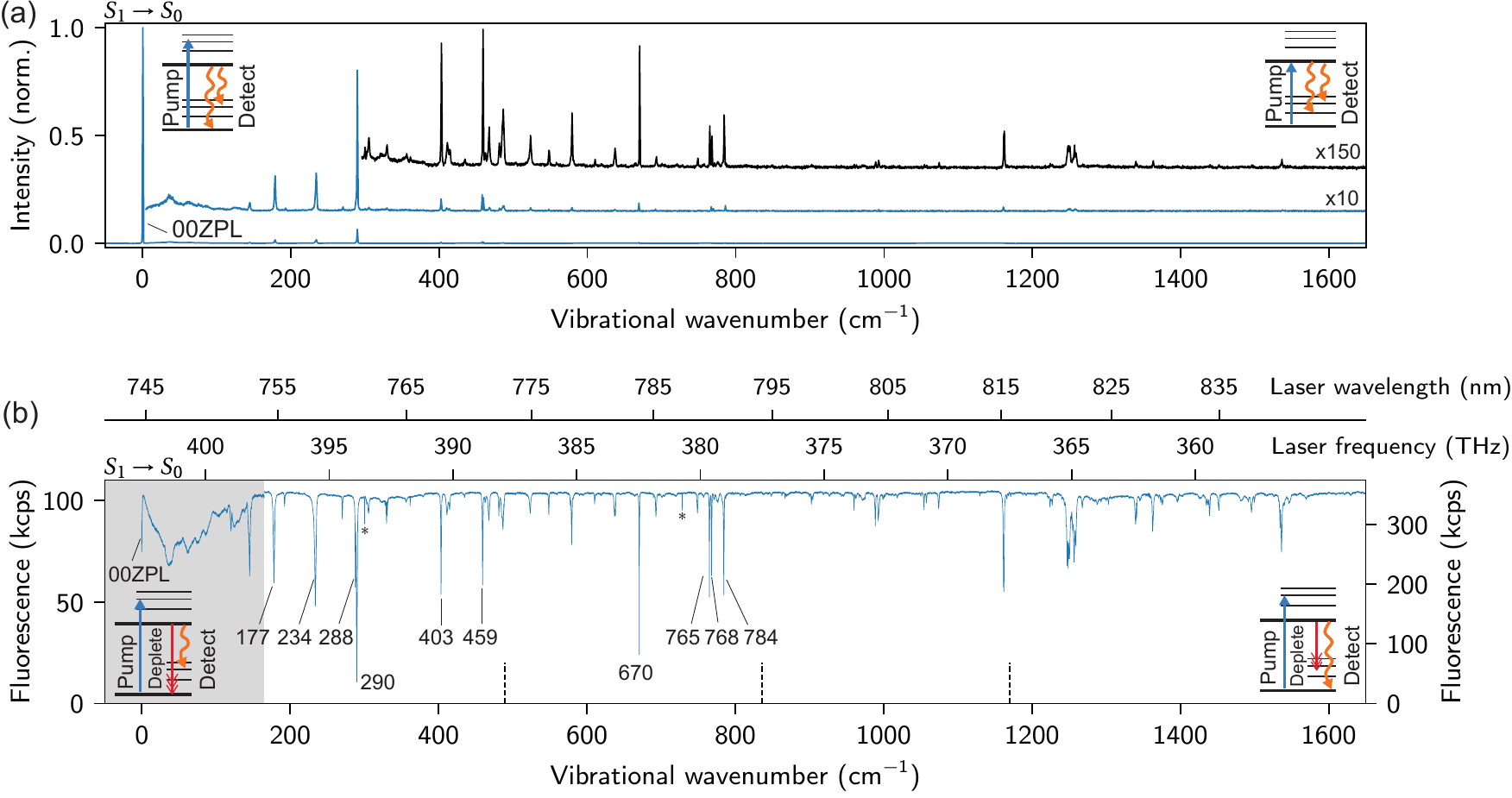}
  \caption{Vibronic spectra of the electronic ground state of a single molecule. (a) Emission recorded with a grating spectrometer. Same as in Figure \ref{spectra_conventional}(b) but extended over a larger spectral range. The blue curves were recorded when exciting the molecule via a vibronic level in $S_1$. The upper curve shows a four-fold magnification. The black spectrum was recorded while exciting the molecule via its 00ZPL and corresponds to a 150-fold magnification of the signal.
 (b) Stimulated emission depletion spectrum on the same molecule as in Figure \ref{spectra_conventional}.
  The phonon wing (gray background, left intensity axis) was recorded using $\Pp =\SI{86}{\micro W}$ and  $\Pd=\SI{501}{\micro W}$.
 For the vibronic part of the spectrum (white background, right intensity axis), 
 we used $P_{\mathrm{p}}=\SI{18}{\micro W}$ ($S_{\mathrm{p}}=0.9$) and $P_{\mathrm{d}}=\SI{102}{\micro W}$.
The black dashed lines at the bottom show the division between the scan regions for which the focus of the laser was optimized separately.
The numbers in the plot mark selected vibronic transitions analyzed in this work.
The regions with narrow vibronic states shown in Figure \ref{narrow_lines} are indicated by an asterisk.}
  \label{spectrum_sep}
\end{figure*}
The vibronic spectrum of $\lvert S_1, \vec{0} \rangle \rightarrow \lvert S_0, \vec{v} \rangle$ transitions is usually obtained by using grating spectrometers \cite{tchenio-1993, tchenio-1993-2, tchenio-1993-3, myers-1994, fleury-1995, jelezko-1996, boiron-1996,  kummer-1997, kulzer-1997, bach-2000, kiraz-2003, lettow-2007, nicolet-2007, trebbia-2009, verhart-2016, tuerschmann-2017}. In Figures \ref{spectra_conventional}(b) and \ref{spectrum_sep}(a), we display the fluorescence emission spectrum of a single molecule that was excited via a $\lvert S_0, \vec{0} \rangle \rightarrow \lvert S_1, \vec{w} \rangle$ transition (and via the 00ZPL for the $\times 150$ version). Recording the red-shifted fluorescence on $\lvert S_1, \vec{0} \rangle \rightarrow \lvert S_0, \vec{v} \rangle$ with a grating spectrometer clearly reveals the 00ZPL, the phonon wing and many vibronic features in the range up to \SI{1650}{cm^{-1}}. This spectrum is also affected by chromatic aberrations induced by the aspheric lens as well as the variation in the quantum efficiency of the spectrometer camera, effectively reducing the signal of vibronic transitions to modes with high wavenumbers. Thus, it should be kept in mind that the relative sizes of the vibronic features in the emission spectrum shown here are not good measures for their relative FC factors. In the fluorescence excitation spectra (see Figure \ref{spectra_conventional}(a)), we reduced the effect of chromatic aberrations by refocusing at regular intervals.  However, the relative intensities of various spectral components are affected by different degrees of saturation. Since high-resolution measurements with grating spectrometers become prohibitively difficult for linewidths below about 10\,GHz, they are not well-suited for a search for narrow vibronic states in $S_0$. Narrow low-amplitude features in the spectrum might even be obscured in such spectra because their intensity is distributed over several noisy camera pixels.

\subsection*{Stimulated emission depletion spectroscopy of vibrational levels in $S_0$}
We now present a method to investigate $\lvert S_1, \vec{0} \rangle \rightarrow \lvert S_0, \vec{v} \rangle$ transitions at high spectral resolution. We start by populating the electronic excited state via the most prominent vibronic transition at \SI{291}{cm^{-1}}. This state quickly relaxes to the vibrational ground state $\vert S_1, \vec{0} \rangle$, which can in turn decay radiatively via the manifold of the vibrational levels in the electronic ground state. Here, we scan the frequency of a second laser beam to stimulate emission to various $\vert S_0, \vec{v} \rangle$ levels, while detecting the 00ZPL fluorescence signal (see insets in Figure \ref{spectrum_sep} (a)). Again, depending on the filter setting, we can examine the phonon wing of the 00ZPL or transitions to vibronic states in $S_0$. Figure \ref{spectrum_sep}(b) displays a remarkable series of high-resolution dips that mark vibronic transitions. The linewidth of the depletion laser used in our experiments is $<\SI{1}{MHz}$ and is monitored at an absolute(relative) accuracy of 60(2)\,MHz. 
Thus, the resolution of our method for determining the linewidths of vibronic transitions is more than four orders of magnitude better than that of the best grating spectrometers \cite{tu-2021}. Our knowledge of the linewidth of the ground state vibrational levels, however, is limited by the width of the 00ZPL ($\sim\SI{30}{MHz}$; see SM, section VI for more details).

At $T \lesssim \SI{10}{K}$, the levels $\vert S_0, \vec{v} \rangle$ carry negligible thermal population. 
Therefore, population inversion between $\vert S_1, \vec{0} \rangle$ and $\vert S_0, \vec{v} \rangle$ is achieved even at very low pump powers $\Pp$. We performed most experiments at pump saturation parameters $S_{\mathrm{p}} \sim 1$ to limit the required laser power while still reaching a good SNR. Under these conditions, the power of the depletion laser beam ($\Pd$) required to saturate stimulated emission is comparable to the saturation power in fluorescence excitation measurements on $\vert S_0, \vec{0} \rangle \rightarrow \vert S_1, \vec{w} \rangle$. In order to increase the signal from less prominent states in the vibronic spectrum, we generally increased $\Pd$. As indicated in Figure \ref{spectrum_sep}(b), we adapted the experimental arrangement to measure the signal of the phonon wing. Here, we collect the red-shifted emission to the most prominent vibrational state in order to avoid the leakage of the strong depletion laser beam through the bandpass filters as its frequency approaches that of 00ZPL. Additionally, we used higher laser powers for both pump and depletion lasers because of the reduction of the collected signal and the reduced interaction cross section for stimulated emission in this configuration. As for the broadband fluorescence excitation spectra shown above, we compensated for the chromatic aberration by refocusing the depletion laser beam in $\sim\SI{10}{THz}$ intervals.

Stronger depletion dips experience more power broadening than weaker dips. High depletion laser powers additionally lead to the reduction of the `baseline' emission because of stimulated emission to phonon sidebands of vibrational states and a dense bath of weak combination and overtone modes. An example of two STED spectra recorded at different depletion powers is shown in Section VI of the SM. To determine the decay rates of the vibronic states in $S_0$, we account for line broadening and baseline depletion effects by fitting a suitable rate equation model to the data.

We remark that the depletion laser field can not only induce incoherent stimulated emission along the transitions $\vert S_1, \vec{0} \rangle \rightarrow \vert S_0, \vec{v} \rangle$, but it can also lead to coherent (stimulated Raman) interactions if it is resonant with $\vert S_1, \vec{w} \rangle \leftrightarrow \vert S_0, \vec{v} \rangle$ transitions.
For most combinations of vibrational states $\vert \vec{w} \rangle$ and $\vert \vec{v} \rangle$, 
the FC overlap is too low, however, to generate noticeable coherent effects at the power levels used for STED spectroscopy in this study.
However, if a state $\vert v \rangle$ has a a similar vibrational wave function as $\vert w \rangle$,
their FC overlap can be comparable to the one of the 00ZPL.
Transitions between these states can reach cross sections similar to stimulated emission.
The corresponding spectral features appear in the vicinity of the 00ZPL and will be presented in a separate study \cite{zirkelbach-2022}.

\section*{Vibronic spectral inhomogeneities}
\begin{figure*}
  \includegraphics{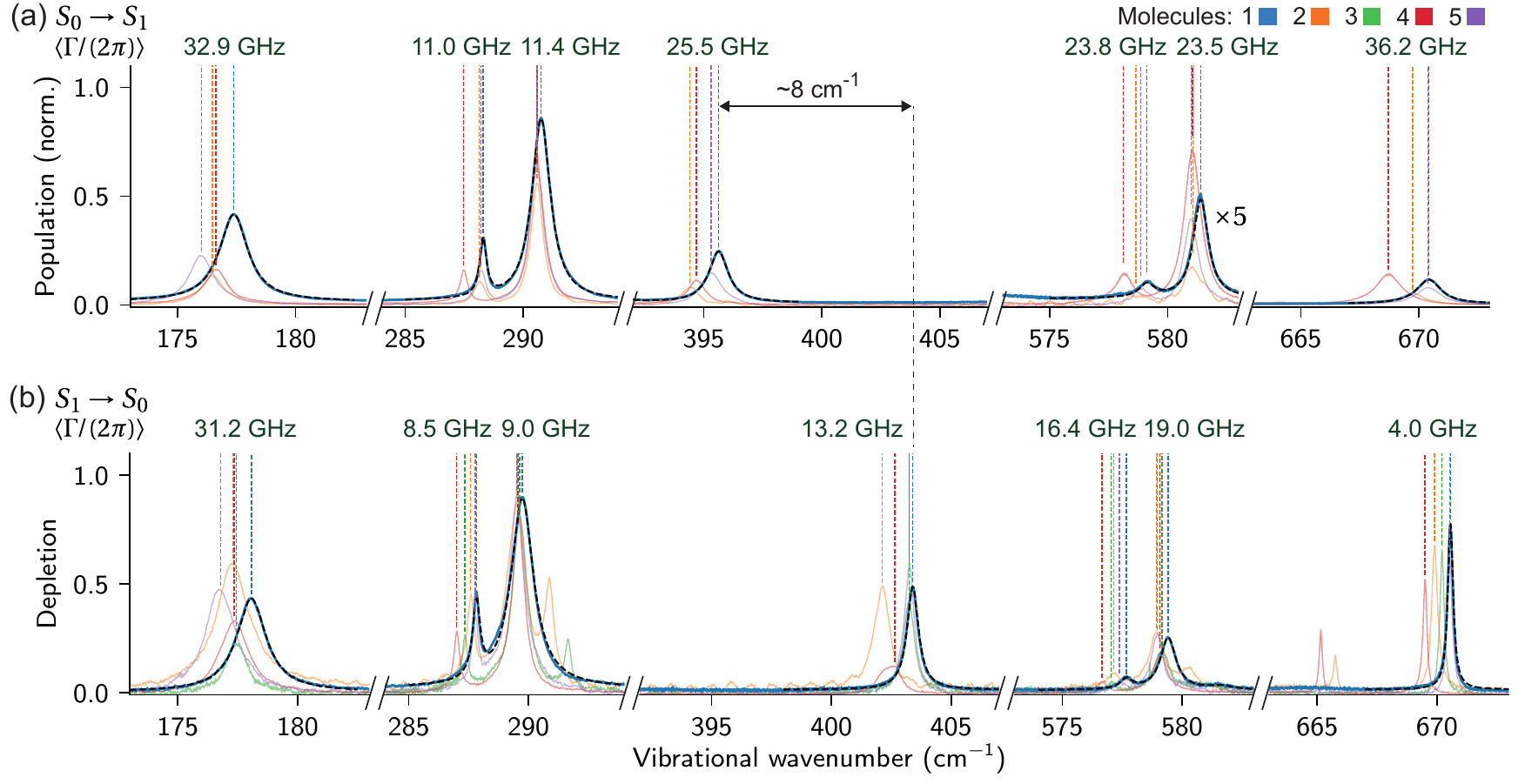}
  \caption{Selected vibronic features in $S_1$ (a) and $S_0$ (b), recorded for four (a) and five (b) different DBT molecules on three samples.
  The black dashed lines show fits to a rate-equation model. The vertical colored dashed lines show the frequency of the peak of each vibronic resonance.
  The average linewidths at low excitation power are indicated on top of the plot for each state ($\Gamma_{we}/(2\pi)$ in (a) and $\Gamma_{vg}/(2\pi)$ in (b), respectively).
  The power of the pump and the depletion lasers used for these experiments vary between the molecules and are listed in Section VII of SM.}
  \label{vibronic_features}
\end{figure*}
The high resolution of our measurements allows us to explore the inhomogeneous distribution of the vibronic transition frequencies and linewidths caused by defects and imperfections in the crystal structure. To have an idea of the typical variations among molecules, we recorded fluorescence excitation spectra of four molecules and STED spectra of five molecules on three crystal samples (see Section VII of SM for the full spectra). Figure \ref{vibronic_features} shows selected regions of these scans for all investigated molecules. For the data from $S_1$, we show the normalized excited state population obtained via normalizing the (background-corrected) APD count rate $R$ by the count rate $R_{\infty}$ at full saturation. The results of the STED measurements are displayed in terms of a depletion factor $D = \bigl[R(\Pd =0) - R\bigr] / R(\Pd = 0)$.\cite{kittrell-1981}

The frequencies of the vibrational states clearly differ for the studied molecules as illustrated in Figure \ref{vibronic_features}. Variations of the linewidth and the FC factors can be difficult to appreciate in a visual inspection because the power and interaction efficiency of the depletion laser vary between the measurements for each molecule. Nevertheless, it is instructive to discuss some of the observed spectral features. 

First, we note the broad level at around $\SI{177}{cm^{-1}}$, which belongs to a state that decays via direct coupling to matrix phonons \cite{hill-1988, dlott-1989}. Next, we point to the most prominent transition in both $S_0$ and $S_1$, namely the peak at around $\SI{290}{cm^{-1}}$. We find a satellite feature at a lower wavenumber for this mode in all our measurements and an additional feature at higher wavenumbers in $S_0$ for two of the five molecules. The peaks at around $\SI{580}{cm^{-1}}$ represent the first overtone of the most prominent level as well as its combination with its satellite level at lower vibrational frequencies. The wavenumbers of both the overtone and the combination mode deviate by less than $\SI{0.15}{cm^{-1}}$ from the sum of the wavenumbers of the corresponding fundamental modes, indicating very little anharmonicity. 

Based on the ratio of the squared Rabi frequencies obtained from the rate equation fit, we estimate $\langle\alpha_i\rangle = \langle\Delta Q_i / (2\Delta Q_{i,\mathrm{zpm}})\rangle = \langle(\Omega_{\mathrm{d},580}^2 / \Omega_{\mathrm{d},290}^2)^{\frac{1}{2}}\rangle = 0.31$ for $S_0$ in the (non-mixing) displaced harmonic oscillator approximation (see Section I in SM), where $\alpha_i$ is the square root of the Huang-Rhys factor\cite{dejong-2015} of mode $i$ and $\Delta Q_{i,\mathrm{zpm}} = (\hbar/(2\omega_i))^{1/2}$. The observation that $\alpha_i$ is considerably smaller than 1, even for the most prominent vibronic mode, shows that the relative displacement of the potential energy surfaces is indeed low in DBT. Figure \ref{vibronic_features} additionally shows that the frequency of the mode at around $\SI{400}{cm^{-1}}$ changes by $\sim\SI{8}{cm^{-1}}$ between $S_0$ and $S_1$, which is considerably more than the observed average frequency difference. Another noteworthy effect is encountered in the mode at $\SI{670}{cm^{-1}}$ which has about 10 times larger linewidth in $S_1$ than in $S_0$. 

We point out that a small fraction of the resonances show asymmetric line shapes. An example is presented in Figures \ref{spectra_conventional} and \ref{spectrum_sep} by the vibrational state around 234\,$\rm cm^{-1}$.\cite{verhart-2016}  One possible cause for the asymmetry of vibrational line shapes is a strong gradient of the phonon density of states over the frequency range of a vibrational state\cite{cusco-2007, reynolds-1972}. A detailed analysis of this phenomenon is beyond the scope of our current study.

\begin{figure*}
  \includegraphics{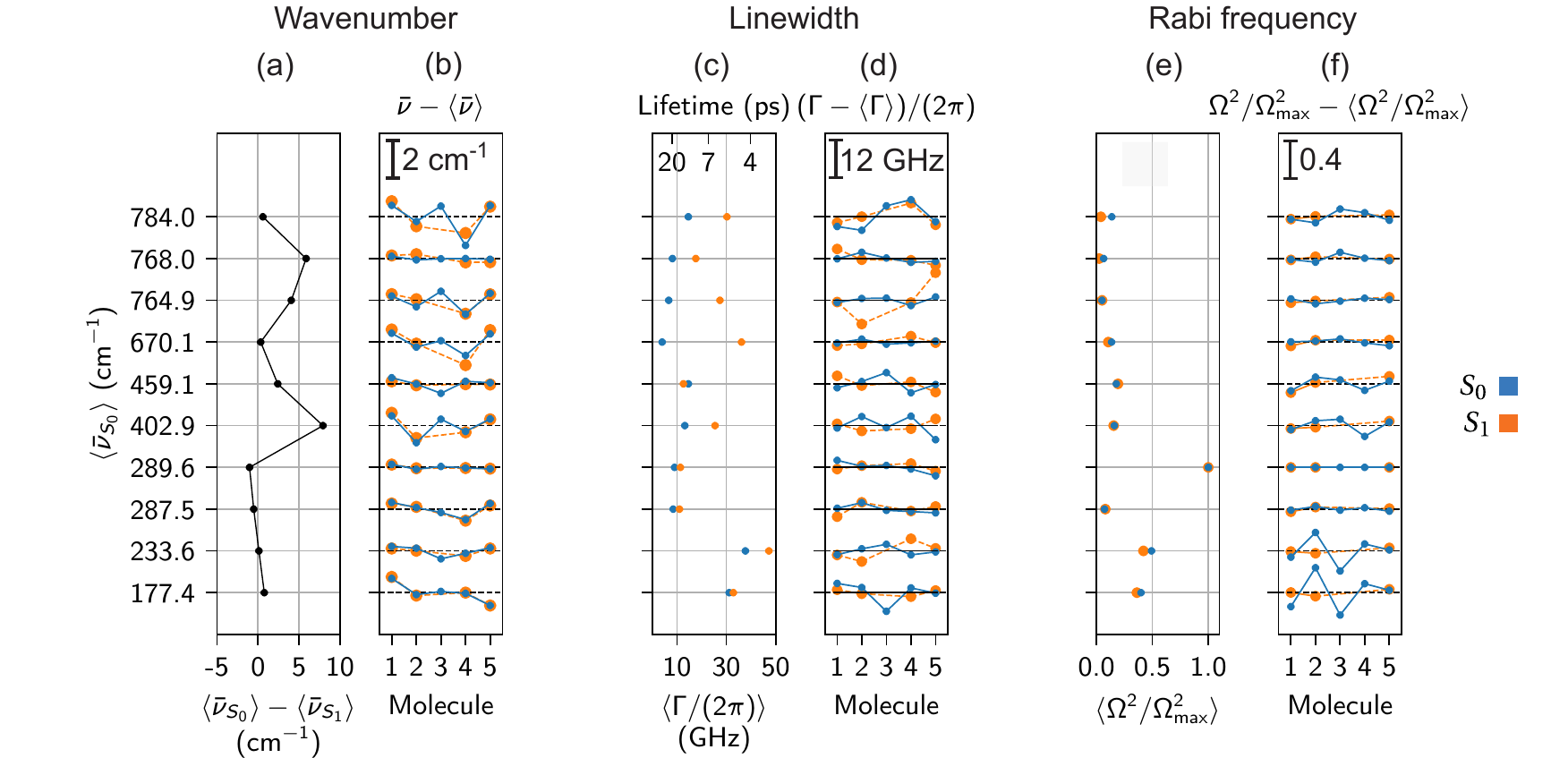}
  \caption{Summary of data on vibronic states for several molecules.
  (a) Differences between the average wavenumbers of the selected modes in $S_0$ and $S_1$.
   The labels of the $y$-axis indicate average wavenumbers in $S_0$.
  (b) Variation of the wavenumbers around their average value. 
  (c) Average linewidths of the selected modes in $S_0$ and $S_1$.  
  (d) Deviation of the vibronic linewidths from their average values.
  (e) Average relative squared Rabi frequencies of the selected modes in $S_0$ and $S_1$.
  (f) Deviations of the relative squared Rabi frequencies from their average values.
  Assuming no change in excitation efficiency between the vibronic modes, these values are proportional to the Franck Condon factors.
  The bars in (b,d,f) indicate the scale of the deviations for each model parameter.}
  \label{stats}
\end{figure*}
To determine the frequencies, linewidths and relative squared Rabi frequencies for a number of selected vibronic features, we fitted them with a model for the excited state population (see the black dashed lines in Figure \ref{vibronic_features} and Sections V and VI in SM for details). The results of these fits are shown for $S_0$ (blue) and $S_1$ (orange) in Figure \ref{stats} for the vibrational modes of each investigated molecule marked in Figures \ref{spectra_conventional}(a) and \ref{spectrum_sep}(b). Fluorescence excitation spectra are available only for molecules 1, 2, 4, and 5, respectively. The data from molecule 4 are omitted in Figure \ref{stats}(e,f) because its fluorescence excitation spectrum showed clear signs of uncorrected chromatic aberrations.

Figure \ref{stats}(a) reveals that the mode frequencies tend to be higher in $S_0$ than in $S_1$. This trend is expected since the strength of the bonds decreases when an electron is promoted to the excited state. The range of wavenumber variation between the molecules shown in Figure \ref{stats}(b) suggests that some levels (for example the levels at $\SI{290}{cm^{-1}}$ and $\SI{768}{cm^{-1}}$ in $S_0$) are less affected by the pDCB crystal environment than others. The average variation of the vibrational frequencies over the selected modes amounts to $\SI{0.9}{cm^{-1}}$ in both $S_0$ and $S_1$.

Figure \ref{stats}(c, d) shows that the vibrational linewidths vary by as much as about one order of magnitude for the selected states. In $S_0$, the linewidths of the states with the lowest frequencies are significantly higher than the other linewidths. This observation is in accordance with the different regimes of vibrational relaxation suggested by the studies of Dlott and coworkers \cite{hill-1988, hill-1988-2}: vibrational states in regime I with wavenumbers below the two-phonon cut-off frequency are expected to decay faster than regime II states above that frequency because they can directly excite pairs of matrix phonons. The narrowest observed linewidth among the selected prominent vibronic states amounts to $\SI{4}{GHz}$ for the level at $\SI{670}{cm^{-1}}$ in $S_0$. Similar to the variation of the vibrational frequencies between individual molecules, the linewidths of vibronic states also show an intermolecular distribution (see Figure \ref{stats}(c)). The linewidths for the selected vibronic levels vary by about \SI{2.4}{GHz} among different molecules in both $S_0$ and $S_1$.

Another interesting observation is that the linewidths of vibronic states in $S_1$ are, on average, significantly higher than for the corresponding modes in $S_0$. For the mode at $\SI{670}{cm^{-1}}$, we even find an almost 10-fold increase in $S_1$. In the context of cubic anharmonic coupling theory for vibrational relaxation \cite{califano-1981, dlott-1989, hill-1988}, the decay rate of a vibrational state depends on the two-phonon density of states and the anharmonic coupling strength of a mode to the matrix.
Since the wavenumber of the mode at $\SI{670}{cm^{-1}}$ does not considerably change between $S_0$ and $S_1$, the observed linewidth difference cannot be explained by a change of the density of states. A larger coupling of DBT in $S_1$ state to the lattice degrees of freedom of the pDCB crystal, however, might be at work.
We are not aware of many studies comparing the linewidths of vibronic states in $S_0$ and $S_1$.\cite{hill-1988-2}
In one example, a 3.5-fold increase of vibrational relaxation in $S_1$ has been reported for pentacene in benzoic acid.\cite{carlson-1990}
In this case, the authors tentatively suggest phonon-assisted decay via a short-lived librational phonon in $S_1$.
Hence, the linewidth differences reported in Figure \ref{stats}(c) prompt and motivate future theoretical efforts to understand vibrational decay dynamics in molecular crystals.

The variation of the amplitudes among vibronic features in the spectra of Figures \ref{spectra_conventional}(a), \ref{spectrum_sep} and \ref{vibronic_features} stems from the differences of the FC factors of the vibronic transitions and their linewidths although it could also be affected by the chromatic aberrations in our optical setup. Under the assumption of constant laser intensity at the position of the molecule, the (relative) squared Rabi frequencies $\Omega^2$ obtained from our model fit to the data are expected to be proportional to the (relative) FC factors of the transitions. The consistent behavior of the average squared Rabi frequencies $\langle \Omega^2 / \Omega_{\mathrm{max}}^2\rangle$ in $S_0$ and $S_1$ shown in Figure \ref{stats}(f) suggests that chromatic effects have been well compensated for by our re-alignment procedure, allowing us to interpret the results as FC factors in a tentative manner. 

\begin{figure}
  \includegraphics{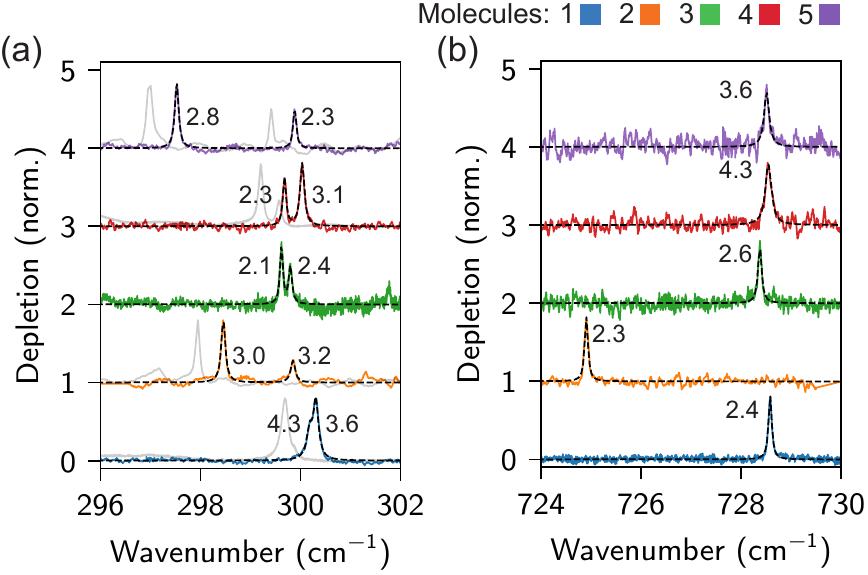}
  \caption{Examples of narrow vibronic lines of single DBT molecules in a pDCB crystal.
  (a) Vibronic lines in the spectral region around $\SI{300}{cm^{-1}}$ in $S_0$.
  The gray lines show the corresponding spectral regions in $S_1$.
  (b) Vibronic lines in the spectral region around $\SI{725}{cm^{-1}}$ in $S_0$.
The black dashed lines show fits of Lorentzian line profiles.
The numbers indicate the linewidth of the fitted Lorentzian function in units of \si{GHz}.}
    \label{narrow_lines}
\end{figure}
In some spectral regions of the vibronic spectra we encountered particularly narrow lines, both in $S_0$ and $S_1$. Figure\,\ref{narrow_lines} shows two of these regions. The linewidth of some of these features is as low as \SI{2}{GHz}, corresponding to a vibrational lifetime of $\SI{80}{ps}$. This is significantly longer than the average lifetime around \SI{15}{ps} for  the prominent modes in $S_0$ and \SI{8}{ps} for the prominent modes in $S_1$. Interestingly, we also found similarly narrow linewidths in $S_1$ (grey spectra) and $S_0$ (colored spectra) in the $\SI{300}{cm^{-1}}$ region (see Figure\,\ref{narrow_lines}(a)), but the narrow lines observed in the region around $\SI{728}{cm^{-1}}$ did not have a counterpart in $S_1$ (see Figure\,\ref{narrow_lines}(b)).

\begin{figure*}
  \includegraphics{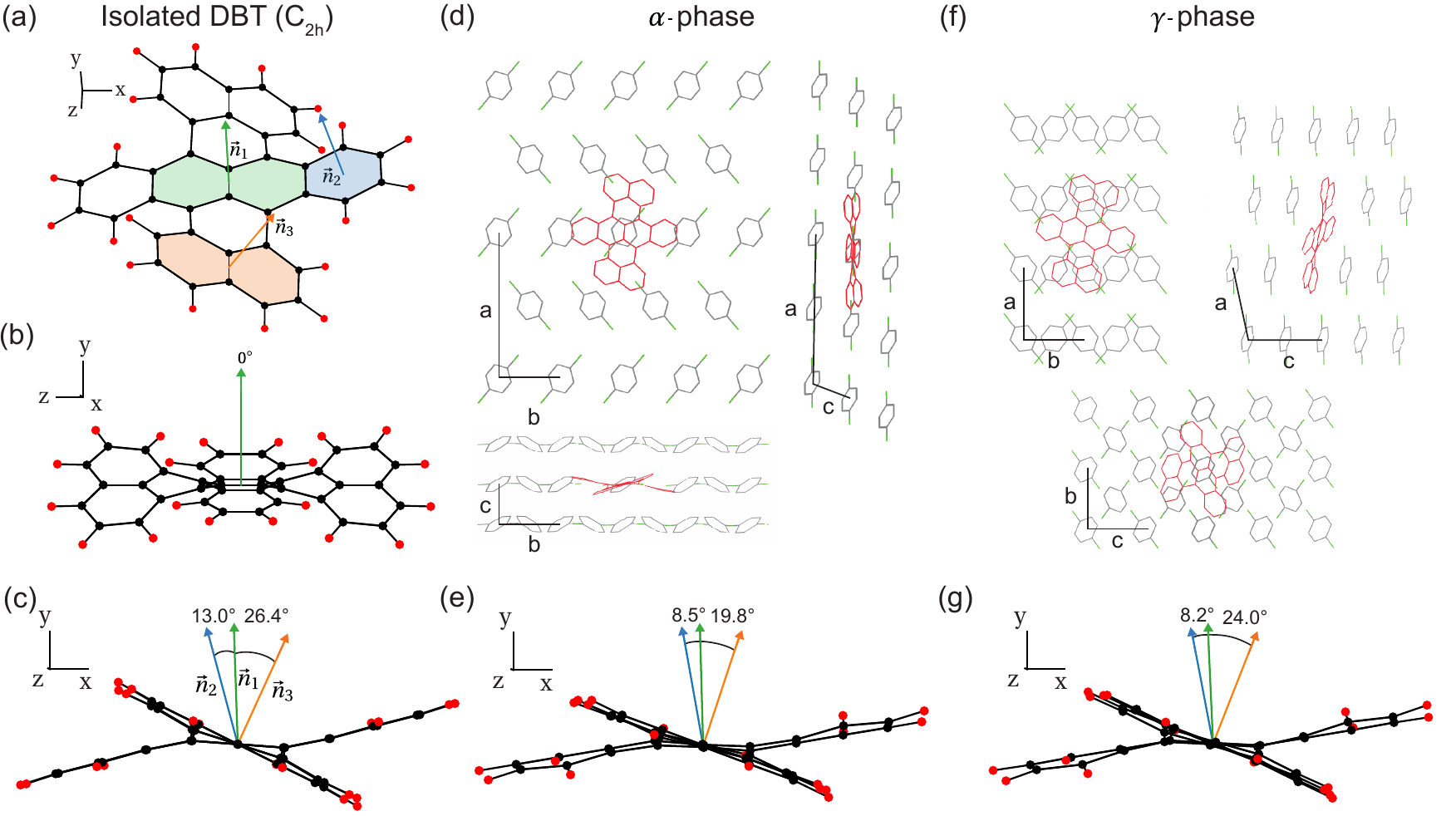}
  \caption{Spatial conformation of DBT resulting from DFT calculations.
  (a) Most stable stereoisomer of isolated DBT. Several planes of the molecule and their normal vectors are highlighted. The $z$-axis of the coordinate system is aligned with the long axis of the terrylene moiety.
  (b,c) Different perspectives of the structure shown in (a) and the angles between the normal vectors defined in (a).
  (d,f) Optimized structures of DBT (red) in $\alpha$- and $\gamma$-pDCB (black), respectively, using the ONIOM method. Hydrogen atoms are not shown.
  The sketched embedding of DBT in the crystal structures marks only one of many possible options.
  (e,g) Equivalent perspective as in (c) but simulated for a molecule embedded in $\alpha$- (e) and $\gamma$-pDCB (g) phases of pDCB.}
  \label{dft} 
\end{figure*}

\section*{Comparison with theoretical predictions}
Having presented experimental vibronic spectra of single molecules in the solid state with unprecedented high resolution, we now explore the agreement of our data with theoretical expectations.
The spatial conformation of the most stable isomer of isolated DBT determined from density functional theory (DFT) calculations is shown in Figure \ref{dft}(a-c). 
When embedded in a pDCB crystal, the conformation of DBT is expected to change, thus, affecting the intensity distribution of its vibronic spectra. To take into account the effect of the crystal on the geometry of the DBT molecule, we employed the ONIOM method with DBT on the B3LYP/6-31G(d,p) level and the rigid pDCB crystal on the PM3 level.\cite{deperasinska-2017, bialkowska-2017}
Since we neither collected structural nor directional information about pDCB at the position of the investigated molecules,
the crystal phase and the alignment of the DBT with respect to its axes are unknown to us.
We, thus, explored the influence of possible DBT embeddings within the $\alpha$- and $\gamma$-phases of pDCB (see Figure \ref{dft}(d,f)).
For $\alpha$-pDCB\cite{estop-1997}($\gamma$-pDCB\cite{wheeler-1976}), we replaced seven(six) crystal molecules by a single DBT molecule and included 62(44) pDCB molecules in the simulations. The crystal structures and optimized geometries of DBT are shown in Figure \ref{dft}(d,f).
To demonstrate the effect of compression exerted by the crystal, we show the deformed DBT molecules from the same perspective as the isolated version in Figure \ref{dft}(e,g).
The presence of the crystal molecules leads to smaller angles between the outer rings and the central plane of DBT. 
Additionally, the crystal induces a twist of the molecular planes, reducing the symmetry of the molecule to the $C_i$ point group. 
As shown in previous studies, the loss of symmetry elements can lead to the activation of additional lines in the vibronic spectrum.\cite{deperasinska-2017, bialkowska-2017}

\begin{figure*}
  \includegraphics{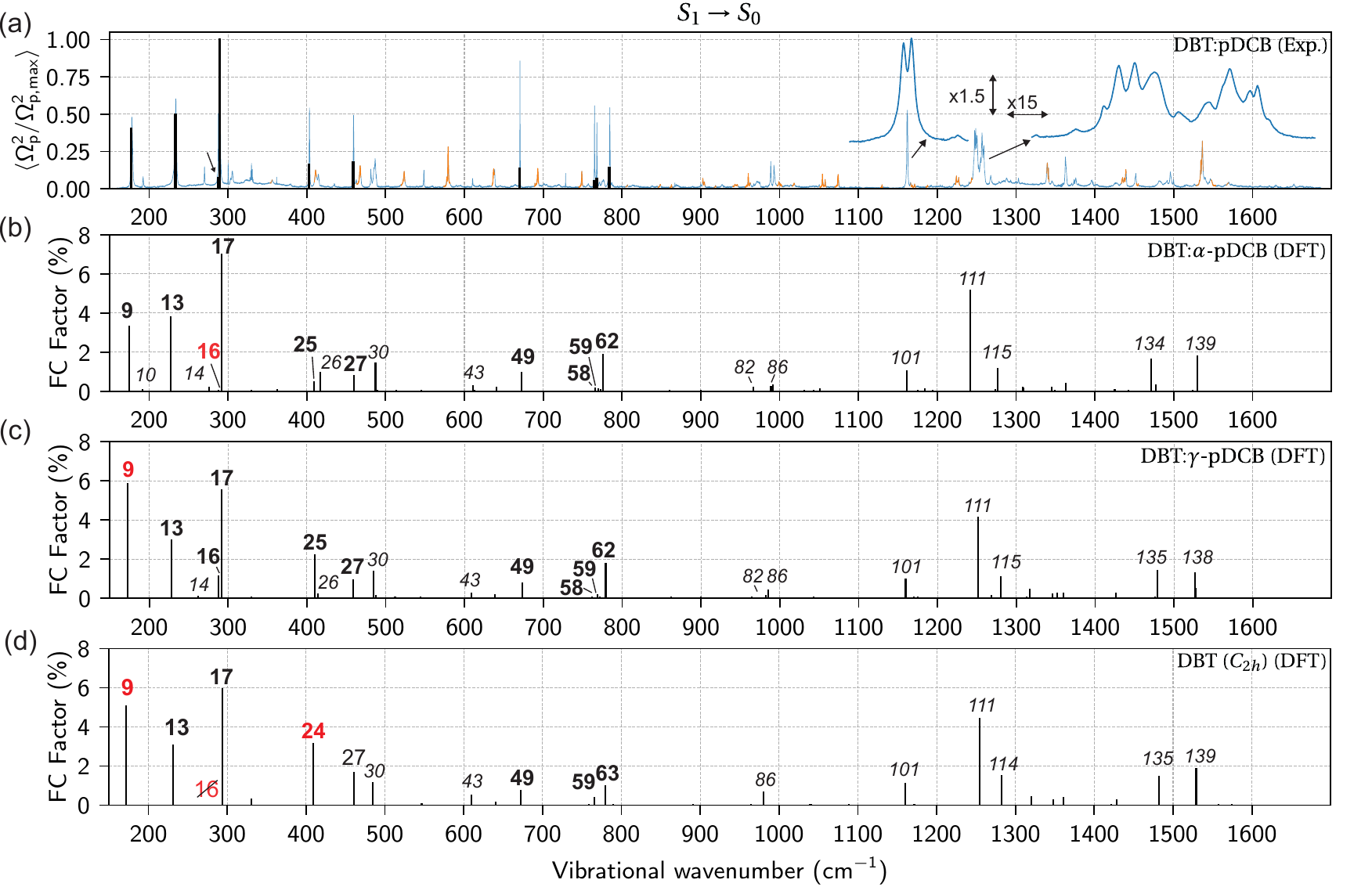}
  \caption{Experimental versus simulated vibronic spectra of vibronic transitions from $S_1$ to $S_0$ in DBT.
  (a) Black lines: Relative squared Rabi frequencies of the selected modes indicated in Figure \ref{spectrum_sep}(a), averaged over five DBT molecules on three samples. 
  The arrow indicates the result for the small satellite peak next to the most intense line.
  Blue line: stimulated emission depletion spectrum from Figure \ref{spectrum_sep}(b).
  Magnified versions of the regions between 1155--$\SI{1168}{cm^{-1}}$ and 1240--$\SI{1264}{cm^{-1}}$ are displayed with an offset of $0.3$.
  Orange lines: regions in the spectrum in which overtone and combination modes of the prominent peaks are expected.
  (b-d) Calculated vibronic spectra ($S_1 \rightarrow S_0$) of DBT in $\alpha$-pDCB, $\gamma$-pDCB, and vacuum, respectively.
  Only fundamental modes are displayed.
  The numbers in the plot are the mode IDs, defined by increasing frequency for each calculation.
  Red numbers indicate considerable deviation with respect to the experimental findings.
  Italic numbers indicate modes that have not been analyzed in more detail in this study.
  The wavenumbers of the calculated modes have been corrected by the linear scaling functions given in the SM, section IX.}
  \label{dft_spectra} 
\end{figure*}

The blue curve in Figure \ref{dft_spectra}(a) shows the experimentally recorded vibronic spectrum of DBT:pDCB in $S_0$, and the orange parts point to regions in which overtone and combination modes of the prominent peaks are expected. We also mark the average line intensities $\langle \Omega_{\mathrm{p}}^2 / \Omega_{\mathrm{p, max}}^2 \rangle$ obtained for various molecules by the black lines. In Figure \ref{dft_spectra}(b-d), we display the results of DFT calculations for transitions $S_1 \rightarrow S_0$ under three conditions of $\alpha$-pDCB, $\gamma$-pDCB and vacuum. In the calculated spectra, we only present the fundamental vibronic modes for better readability. Their mode numbers are indicated next to each peak. Red numbers indicate cases for which the calculated mode intensity differs considerably from the experimental observations.

The theoretical predictions for the vibronic spectra of isolated DBT (see Figure \ref{dft_spectra}(d)) do not agree well with our STED spectroscopy results.
The calculated intensities of modes 9 and 24 are higher than our experimental findings.
Moreover, while mode 16 of isolated DBT would have a suitable frequency to explain the satellite peak next to the most intense line, 
this mode is not active due to its B$_{\mathrm{g}}$-symmetry.
Reduction of the molecular symmetry causes the spectra of the deformed DBT molecules to contain more lines than in isolated DBT.
We find that mode 16 becomes active in calculations for both $\alpha$- and $\gamma$-phase pDCB. 
Moreover, the intensity of mode 24 from isolated DBT redistributes its intensity over modes 25 and 26, yielding a better agreement between theory and experiment.
Overall, the vibronic spectrum resulting from the $\alpha$-pDCB simulation matches our experimental observations better, mainly because it reproduces the relative intensities between mode 17 and mode 9 more accurately.
We can, however, not exclude that a different embedding of DBT in $\gamma$-pDCB is able to yield an even better agreement with the experimental spectra than the current calculations for $\alpha$-pDCB. Bulk pDCB crystals at liquid Helium temperatures are rather expected to be found the $\gamma$-phase if they are cooled down slowly.\cite{zikumaru-1988, ghelfenstein-1971, jongenelis-1989, verhart-2016} In our experiments, we cooled pDCB crystals enclosed in nanochannels from room temperature to liquid helium temperature at pressures $<10^{-3}\si{mbar}$. While we are not certain in which phase the pDCB crystal forms under these conditions, we find that besides the prominent features analyzed in the experimental data, there also exist less intense peaks that are predicted by the simulations (e.g.~the modes 10, 14, 43, and the modes around 86). On the other hand, we also observe spectral features that are not predicted by the calculations. In particular, the narrowest resonances at $\SI{300}{cm^{-1}}$ and $\SI{725}{cm^{-1}}$ with linewidths about 2\,GHz are not mirrored in the calculations.

\begin{figure*}
  \includegraphics{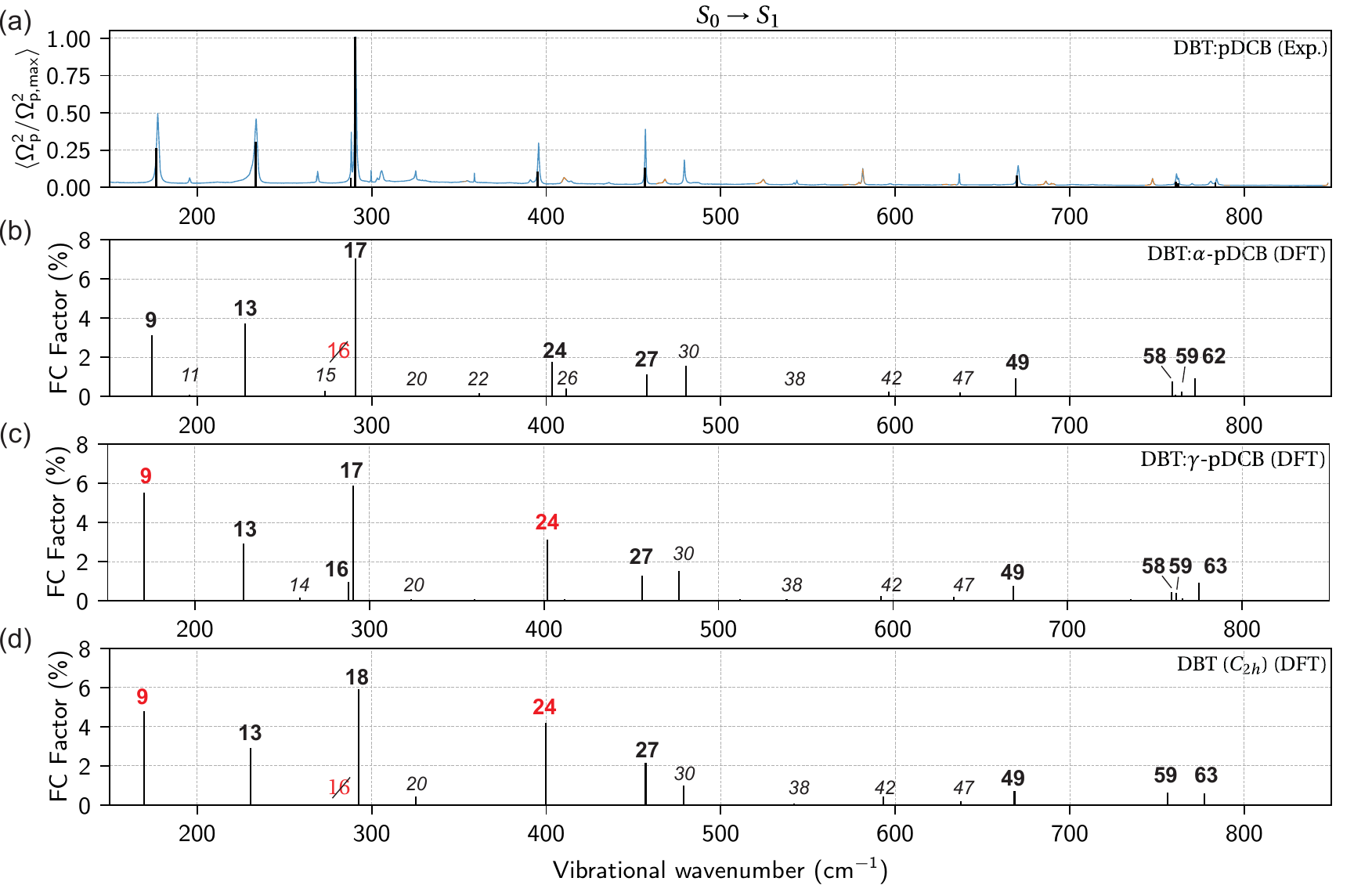}
  \caption{Experimental versus simulated vibronic spectra of vibronic transitions from $S_0$ to $S_1$ in DBT.
  (a) Black lines: Relative squared Rabi frequencies of the selected modes indicated in Figure \ref{spectra_conventional}(a), averaged over three DBT molecules on three samples. 
  Blue line: fluorescence excitation spectrum from Figure \ref{spectra_conventional}(a).
    Orange lines: regions in the spectrum in which overtone and combination modes of the prominent peaks are expected.
  (b,c,d) Calculated vibronic spectra ($S_0 \rightarrow S_1$) of DBT in $\alpha$-pDCB, $\gamma$-pDCB, and vacuum, respectively.
  See the caption of figure \ref{dft_spectra} for more details.}
  \label{dft_spectra_s1} 
\end{figure*}

Figure \ref{dft_spectra_s1} shows DFT results for vibronic transitions $S_0 \rightarrow S_1$ and data obtained from our fluorescence excitation measurements. The results of the DFT calculations and their agreement with the data is similar to the results shown in Figure \ref{dft_spectra}. In contrast to the findings for transitions $S_1 \rightarrow S_0$, mode 16 is predicted in the $\gamma$-pDCB but it is not active in $\alpha$-pDCB (see Figure \ref{dft_spectra_s1}(b)). However, for the other prominent modes, in particular modes 9 and 24, the experimental findings match best the calculations for $\alpha$-pDCB. The DFT calculations also predict the large change of the vibrational frequency of the mode around $\SI{400}{cm^{-1}}$ observed in the experiment (see Figure \ref{vibronic_features}).
A complete list of mode assignments to the prominent modes analyzed in Figure \ref{stats} is provided in Section IX of SM.

\section*{Discussion}
The vibronic states of a molecule provide a wealth of information about fine details of the embedding conformation of the molecule and its surrounding. This has already been recognized in the early studies of single molecules where differences between the vibronic spectra of individual molecules were reported. \cite{myers-1994, kummer-1997} In this work, we have set the ground for more sensitive and quantitative studies of the vibrational degrees of freedom of molecules embedded in a solid matrix. We employed fluorescence excitation spectroscopy for probing the vibrational levels of the electronic excited state of single DBT molecules embedded in a pDCB crystal. Furthermore, we introduced STED spectroscopy for accessing the vibrational levels of the electronic ground state, which allowed us to identify vibronic resonances as narrow as 2\,GHz. 
Our method can resolve vibronic transitions at a resolution that is only limited by the linewidth of the laser ($<$ \SI{1}{MHz} in our case), which is four orders of magnitude higher than that of grating spectrometers.
In this approach, one achieves a much higher degree of frequency calibration than in grating spectroscopy since one relies on the fine frequency scan of a  laser.

Developments of computational methods and the rapid increase in their capacities enable one to simulate how dye molecules like DBT are embedded in molecular crystals on a quantum mechanical level. Hence, state of the art quantum chemical simulations can explain the appearance of new lines related to symmetry breaking induced by the crystal matrix.\cite{bialkowska-2017, deperasinska-2017}, although the distribution of the intensities and frequencies of the simulated vibronic lines do not fully agree with the experimental data. This is not very surprising since our treatment did not apply the full quantum mechanical model to the complete crystal system, but it only considered the guest molecule. 
In addition, accurate values of the simulated wavenumbers require mode-specific correction terms. 
Besides conformational changes, one might consider a redistribution of intensities between vibronic lines caused by $^{13}$C substitution of individual carbon atoms in DBT and by Herzberg-Teller effects \cite{myers-1994, herzberg-1933, doppagne-2017, kong-2021}. The probability of $^{13}$C substitution is about 28\% in a DBT molecule and the effect on the vibronic spectrum depends on the exact substitution site. Since the effect of $^{13}$C substitution of a single carbon atom can be regarded as a perturbation of the spectrum of isotopically pure DBT, isotope effects may at most explain small details of the intermolecular variation in our data. For the behavior of the average line properties, however, isotope substitution should not be relevant. As for Herzberg-Teller effects, DFT simulations predict corrections on the few percent level compared to the strongest vibronic features in the FC approximation. We, thus, conclude that these phenomena are not dominant in our case.

The high-resolution measurements presented in this study reveal several interesting properties of vibronic features of the complex polyatomic molecule DBT. One observation for almost all examples shown in Figure \ref{vibronic_features} is that the inhomogeneity of the line positions and linewidths, albeit with data from a small number of studied molecules, is small as compared to the changes of these properties between the ground and excited states. A closer look at Figure \ref{stats}(a) suggests that the vibrational frequencies of molecules 2 and 4 tend to lie below the average of all five molecules whereas the values for molecules 1,3, and 5 are situated above the average value. While the number of molecules investigated in this study is too low for a statistically significant statement, this grouping might indicate different nano-environments of the respective molecules.\cite{tchenio-1993-2} Interestingly, the same classification is also apparent in the intensities of the vibronic features at $\SI{177}{cm^{-1}}$ and $\SI{234}{cm^{-1}}$ in $S_0$. Since our DFT calculations suggest that the modes 9 and 13 can exhibit large intensity changes in different embeddings, they might indeed be useful as sensitive nanoscopic probes for the environment.

Another noteworthy observation points to the line doublet at $\SI{1162}{cm^{-1}}$, measured via STED spectroscopy (see Figure \ref{dft_spectra}(a)). DFT spectra in the harmonic approximation predict only a single active mode at this frequency. One explanation for the doublet might be line splitting due to Fermi resonance \cite{fermi-1931, glazunov-1982, ivanov-2013}.
A similar observation holds for the spectral region around $\SI{1250}{cm^{-1}}$, where our measurements reveal a rich line profile while DFT only predicts a single line. This might be caused by Fermi resonance with lower-lying states of DBT, leading to multiple line splittings. Other doublets and complex line profiles occurring in our data might be caused by the same effect.
The measurement methods presented in our work can be used to explore these phenomena in future experiments.

The relaxation rates of vibrational modes in many molecular crystals can be explained well by cubic anharmonic coupling to crystal phonons.\cite{califano-1981, dlott-1989} Such a decay process involves the excitation of two vibrations with lower energies than the energy of the initial vibration, e.g., via two matrix phonons, one matrix phonon and one molecular vibration, or two molecular vibrations.\cite{dlott-1989} The energy of the initial vibration and the maximum phonon frequency in a crystal determine which decay channel is the dominant process. 
Currently, a complete data set for the relaxation rates of the internal modes of DBT is missing in the literature, hindering a comparison between the measured linewidths and the predictions of a theoretical model \cite{hill-1988-4}. 
According to the model of cubic anharmonic coupling, a vibrational state can have a long lifetime if the density of the accepting modes is low and if the anharmonic coupling to them is weak (Fermi's golden rule).  A prominent example for a long-lived vibrational state in a molecular crystal is the \SI{606}{cm^{-1}} mode of pure benzene with a lifetime of \SI{2.65}{ns}.\cite{trout-1984, velsko-1985} As shown in Figure \ref{narrow_lines}, we also observed some states in DBT with exceptionally narrow linewidths. This observation demonstrates that molecules as large as DBT can still have vibrational states which are to some extent shielded from vibrational decay channels.  

Mechanical decoupling of a molecule from its environment should lead to slower vibrational relaxation rates of all states that require matrix phonons to decay. An interesting question is, hence, whether the lifetime of vibrational states of molecules can be extended to a regime relevant for applications in quantum information processing. 
For instance, one could engineer the density of phononic states in the environment around the molecule. Future experimental and theoretical efforts in our group aim to explore such approaches in more detail.



\bibliography{zirkelbach-lib}

\end{document}


\title{Supplementary Material: Non-linear vibrational spectroscopy with a single dye molecule in a solid}

\author{Johannes Zirkelbach}
\affiliation{Max Planck Institute for the Science of Light, D-91058 Erlangen, Germany}
\affiliation{Department of Physics, Friedrich Alexander University Erlangen-Nuremberg, D-91058 Erlangen, Germany}
\author{Masoud Mirzaei}%
\affiliation{Max Planck Institute for the Science of Light, D-91058 Erlangen, Germany}
\affiliation{Department of Physics, Friedrich Alexander University Erlangen-Nuremberg, D-91058 Erlangen, Germany}
\author{Burak Gurlek}%
\affiliation{Max Planck Institute for the Science of Light, D-91058 Erlangen, Germany}
\affiliation{Department of Physics, Friedrich Alexander University Erlangen-Nuremberg, D-91058 Erlangen, Germany}
\author{Irena Deperasin´ska}%
\affiliation{Institute of Physics, Polish Academy of Sciences, Al. Lotniko´w 32/46,
02-668 Warsaw, Poland}
\author{Boleslaw Kozankiewicz}%
\affiliation{Institute of Physics, Polish Academy of Sciences, Al. Lotniko´w 32/46,
02-668 Warsaw, Poland}
\author{Alexey Shkarin}%
\affiliation{Max Planck Institute for the Science of Light, D-91058 Erlangen, Germany}
\author{Tobias Utikal}%
\affiliation{Max Planck Institute for the Science of Light, D-91058 Erlangen, Germany}
\author{Stephan Götzinger}%
\affiliation{Max Planck Institute for the Science of Light, D-91058 Erlangen, Germany}
\affiliation{Department of Physics, Friedrich Alexander University Erlangen-Nuremberg, D-91058 Erlangen, Germany}
\affiliation{Graduate School in Advanced Optical Technologies (SAOT), Friedrich Alexander
University Erlangen-Nuremberg, D-91052 Erlangen, Germany}
\author{Vahid Sandoghdar}%
\affiliation{Max Planck Institute for the Science of Light, D-91058 Erlangen, Germany}
\affiliation{Department of Physics, Friedrich Alexander University Erlangen-Nuremberg, D-91058 Erlangen, Germany}
\email{vahid.sandoghdar@mpl.mpg.de}

\date{\today}

{
\let\clearpage\relax
\maketitle
}

\tableofcontents

\section{Sample preparation}
\textcolor{blue}{Johannes (or Masoud?)}

\section{Experimental setup}
\textcolor{blue}{Johannes (or Masoud?)}
\begin{itemize}
\item sketch
    \item filter edge measurements?
    \item cryostat layout?
\end{itemize}

\clearpage
\section{Molecule characterization}
\begin{figure*}[!ht]
\includegraphics{figures/si_mol_details.png}
\caption{\label{fig:si_mol_details} 
(a) Fluorescence excitation scan of a range of more than $\pm$1\,THz around the 00-ZPL frequency of the studied molecule. 
Due to the low doping level, there are no other molecules in the same excitation volume, eliminating any inhomogeneous broadening.
(b) Fluorescence excitation scan of the 00-ZPL transition in the low excitation regime yielding a Lorentzian profile with lifetime-limited linewidth.
(c) Saturation scan of the 00-ZPL line.
(d) Fluorescence excitation spectrum showing vibrational states of the molecule in its electronic excited state (P=...). 
At $\nu = ... \,$THz, the laser was refocused and re-aligned to correct for chromatic aberrations of the aspheric lens.
(e) Fluorescence excitation scans of the region around the strongest vibrational line recorded at different excitation powers (P=..., P = ...,). 
(f) Saturation scan of the level shown in (e).}
\end{figure*}

\begin{figure*}[!ht]
\includegraphics{figures/si_mol_autocorr.png}
 \caption{\label{fig:si_mol_autocorr}
(a) Lifetime measurement performed with pulsed excitation of the molecule via a vibrational level. 
(b) Autocorrelation data of the fluorescence photons recorded while (cw-)pumping the molecule via a vibrational level (at $\sim$ 290\,1/cm) at different powers (S = ..., from bottom to top). 
Note that the data are shown with a 0.2 offset between each measurement.
Black lines show the expected, detector-corrected $g^{(2)}$-functions for each pump saturation parameter.
(c) Same measurements as in (b) but plotted over a longer range of delays, showing the photon bunching caused by triplet transitions.
Autocorrelation values in the range $\tau \in [-0.3,0.3]\,\mu$s are not shown in this plot for clarity. 
Black lines are fits of a (single-)exponential model to the bunching peaks.
(d) Contrast and decay of the fits to the data shown in (c).}
\end{figure*}

Flow of the text:
\begin{itemize} 
    \setlength\itemsep{-0.3em}
    \item Low doping level allows us to identify spots on the sample with a single molecule in a given excitation volume. See Figure \ref{fig:si_mol_details}(a) for the (non-existing) inhomogeneous broadening around the molecule for which the data are shown in the main text.
    \item Good crystal properties: reaching lifetime-limited linewidth in the low excitation regime; no pure dephasing present. see Figure \ref{fig:si_mol_details}(b).
    After some time of studying the molecule, also at high powers: observed occasional small jumps of the 00-ZPL (sub-linewidth), but no general broadening.
    \item The fact that only a single molecule is present in the sample allows excitation of the molecule via vibrational transitions. One can even record a single-molecule fluorescence excitation spectrum of the vibrational manifold of the excited state (see Figure \ref{fig:si_mol_details}(d)).
    \item By good suppression of background light (from fiber: via shortpass filter; from sample: via narrow BP filter around the 00-ZPL in detection), we can go to quite high saturation parameters (S$\sim$120) while exciting a vibrational state of the molecule, see Figure \ref{fig:si_mol_details}(e,f).
    
    \item The low doping level also allows for pulsed lifetime measurements; no other molecules can be excited by the relatively broad $\sim$2\,ps pulses from a Mira 900 laser in our setup.
    \item A fit to the exponential fluorescence decay yields $\tau\sim6.92\,$ns, in good agreement with the measured low excitation power linewidth; see Figure \ref{fig:si_mol_autocorr}(a).
    \item Performing cw-autocorrelation measurements show the single molecule nature of emission via $g^{(2)}(0)< 0.5$ (see Figure \ref{fig:si_mol_autocorr}(b)).
    \item Figure \ref{fig:si_mol_autocorr}(b) shows autocorrelation measurements recorded at different excitation powers while the laser was resonant with the strongest vibration at around 290.5\,1/cm. 
    Black lines show the detector-corrected auto-correlation function for this situation \cite{trebbia2009efficient}:
    \begin{equation} \label{Equ:G201D}
        g^{(2)}_d(\tau) = \int_{-\infty}^{\infty}g^{(2)}(\tau - \tau') \cdot \mathrm{IRF}(\tau') \mathrm{d}\tau',
    \end{equation}
    with 
    \begin{equation}
        g^{(2)}(\tau) = 1 - e^{- \Gamma_1 (1 + S) |\tau|}.
    \end{equation}
    The good agreement between data and the model (without fit parameters) shows that poissonian background is negligible for these measurements.
    The fact that $g^{(2)}(0)$ does not get closer to 0 is caused by the two-detector IRF of the HBT-setup which was used for these measurements.
    
    \item Figure \ref{fig:si_mol_autocorr}(c) shows the same measurements as shown in (b), but over a longer time scale (and longer temporal bins).
    We attribute the exponentially decaying feature to intersystem crossing events of the molecule.
    As expected, the contrast of the related bunching feature in the autocorrelation data increases with incident power, because then a triplet transition is better visible as an interruption of the photon flow.
    \item The data in Figure \ref{fig:si_mol_autocorr}(c) were fitted by a single exponential decay model \cite{bernard1993photon, boiron1996single}:
    \begin{equation}
        g^{(2)}(\tau) = 1 + C \cdot e^{-\lambda \vert \tau \vert}.
    \end{equation}
    The contrast and decay rate values obtained from the fits are plotted in \ref{fig:si_mol_autocorr}(d,e).
    The low contrast of the triplet-related bunching is a clear indication of a low triplet yield, which has been reported for DBT:pDCB as well as in other matrices (Anthracene and Naphthalene) before \cite{verhart2016spectroscopy, nicolet2007single,jelezko1996dibenzoterrylene}.
    
    \item \textcolor{violet}{To be checked by Burak} \\
    One can analyze our bunching data in the framework of a mono-exponential decay in more detail (see \cite{bernard1993photon}).
    Both, the contrast $C$ of the bunching peak as well as its decay parameter $\lambda$ are expected to follow a saturation behavior:
    \begin{align}
        \lambda &= \gamma_{31} + \gamma_{31} S / (1+2 S \gamma_{31} / \gamma_{23}) \label{equ:lambda}\\
        C &= (\lambda - \gamma_{31}) / \gamma_{31} \label{equ:contrast},
    \end{align}
    where $S$ is the saturation parameter, \
    $\gamma_{23}$ is the ISC rate into the triplet state,
    and $\gamma_{31}$ is the triplet decay rate.
    As $S$ increases, $\lambda$ changes from $\gamma_{31}$ to $\gamma_{31} + \gamma_{32}/2$ and $C$ changes from 0 to $\gamma_{32} / (2\gamma_{31})$.
    Note that if $\gamma_{23}$ is small, the (relative) change of $\lambda$ can be very small as $S$ is varied.
    Solving equations for $\lambda$ and $C$ in the limit $S\rightarrow\infty$,
    we obtain:
    \begin{align}
        \gamma_{31} &= \lambda / (1 + C)\\
        \gamma_{23} &= 2 \lambda C / (1+C).
    \end{align}    
    
    Our data in Figure \ref{fig:si_mol_autocorr}(c) let us extract the contrasts quite reliably. The fit of a saturation curve as shown in Figure \ref{fig:si_mol_autocorr}(d) results in a maximum contrast of $C = 0.36\%$, yielding $\gamma_{23}/\gamma_{31} = 2 C = 7.2 \cdot 10^{-3}$.
    Since the ratio of transitions into the triplet state vs. transitions out of the triplet state is relevant for the average population trapped in the triplet state, the low observed contrast is equivalent to a low dwell time of the population in the triplet. 
    The fit results for $\lambda$ are not as conclusive as the ones for the contrast.
    Because $\gamma_{32}$ is low compared to $\gamma_{31}$, we expect the saturation curve of $\lambda$ to be almost constant.
    Using the two values at the highest powers to estimate $\lambda$, we obtain $\lambda \approx 117\,$kHz.
    This results in $\gamma_{31}\approx 117\,$kHz and $\gamma_{32} \approx 0.8\,$kHz.
    Note that the triplet state of DBT is expected to have quite low energy and might therefore decay relatively fast because of coupling to vibrational states \cite{jelezko1996dibenzoterrylene}.
    Combining the measured excited state lifetime with the ISC rate results in an ISC yield of about $\gamma_{32} \cdot \tau \approx 5\cdot 10^-6$.
    This is slightly higher than the upper bound of $10^-7$ which has been reported for DBT:Ac \cite{nicolet2007single}.

\end{itemize}

\begin{itemize}
    \item level assignments of identified vibrational states (\textcolor{blue}{Boleslaw and Irena})
    \item pump-dump measurements on 00-ZPL, showing peak to dip transition
\end{itemize}

\clearpage
\section{Theory of pump-dump spectroscopy}
\textcolor{blue}{Johannes, thorough check by Burak}

\begin{itemize}
    \item Depletion process + saturation behavior
    \item Splitting of a vibrational level; should we give a full quantum treatment (\textcolor{blue}{Burak?}) or be happy with coupled oscillators?
\end{itemize}

\section{Data evaluation details}

\clearpage
\section{Saturation behavior of depletion}
\textcolor{blue}{Johannes}
\begin{itemize}
\item Power-broadening correction
    \item \textcolor{red}{missing: high-power pump and dump saturation scans}
\end{itemize}
\begin{figure*}[!ht]
\includegraphics[]{figures/si_pd_broad_sat.png}
\caption{\label{fig:si_pd_broad_sat} 
Pump-dump spectra of a single DBT molecule recorded at two different dump powers. 
For higher dump power, the depletion process is stronger, leading to deeper dips.
Note that both measurements have been done at the same pump power and are plotted without displacement of the baseline.}
\end{figure*}

\begin{figure*}[!ht]
\includegraphics[]{figures/si_dip_saturation.png}
\caption{\label{fig:si_dip_saturation} 
Saturation behavior of the depletion process shown for individual vibrational states.
(a,b) Frequency scans over the resonance of vibrational states at different powers of the dump laser.
The scans are plotted without offset.
(c) Dump laser-induced depletion for the three levels shown in (a,b).
(d) Linewidth of the depletion profile obtained from Lorentzian fits to the data.}
\end{figure*}

\clearpage
\section{Pump-dump vs.~grating spectrometer}
\textcolor{blue}{Johannes}

\section{More details on statistics between molecules / excited vs. ground state}
\textcolor{blue}{Johannes}
\begin{itemize}
    \item \textcolor{blue}{missing: could collect more triplet data from other molecule, but not stricly necessary}
\end{itemize}

\begin{itemize}
    \item Tables showing wavenumbers + linewidths for many identified levels
\end{itemize}
\bibliography{vib_spec_lib} 
\bibliographystyle{apsrev4-1}